# Logistic regression is not enough: The need for Bayesian nonparametric modelling for causal inference using observational data, exemplified by the 'gateway' effect

Floe Foxon, MSc[1,2], Raymond Niaura, PhD[3]

[1] College of Natural Sciences, The University of Texas at Austin, Austin, TX

[2] Department of Data Management & Statistical Analysis, Pinney Associates, Bethesda, MD

[3] School of Global Public Health, New York University, New York, NY

**Corresponding Author** Floe Foxon, ff5384@eid.utexas.edu

## ABSTRACT

**Introduction**: Logistic regression (LR)-type model limitations for causal inference are explained theoretically and empirically through the lens of the purported 'gateway' effect from e-cigarette use to smoking. Previous studies have reported that baseline e-cigarette use quadruples odds of follow-up smoking (binarized) in LR-type models of adolescent longitudinal cohorts (LCs), such that increased e-cigarette use would counteract smoking declines. However, US population-level trends show accelerated smoking declines to record-lows when e-cigarette use increased, presenting an apparent 'paradox'. **Methods**: Population Assessment of Tobacco and Health (USA) Youth Waves 3→4 were analyzed with Bayesian Additive Regression Trees (BART) to model baseline e-cigarette use (treatment) and change in number of days smoking from baseline→follow-up (numerical response) among never- and ever-smoking respondents (group effects), adjusting for confounding risk factors (socio-demographic, intra-individual, behavioural, peer influence, and family background). Unlike LR-type models, BART provides nonlinear, nonparametric modelling with counterfactuals and provides causal effect estimates with principled uncertainty estimation. **Results**: The average effect of e-cigarette use on smoking was both clinically and statistically significant among ever-smoking adolescents (−2 days smoking [diversionary effect; opposite to gateway]) and was not clinically significant among never-smoking adolescents (<1-day absolute change in days smoking [null effect]). **Conclusions**: When LC data are analyzed with causal inference techniques, the gateway effect disappears, consistent with population-level trends. This likely explains why gateway effects predicted in previous LR-type studies have not materialized in a population-level reversal/unexpected slowing of the US adolescent smoking decline, resolving the 'paradox'. Severe LR-type model limitations preclude their use for causal inference.

## INTRODUCTION

Studies using logistic regression (LR)-type linear models applied to observational data are ubiquitous despite severe methodological limitations which preclude causal inference. This is exemplified by the purported 'gateway' effect from e-cigarette use to smoking. We describe the theoretical limitations of LR-type models in detail, and show empirically that when proper causal inference techniques are applied to the same data, the purported effect disappears.

*Evidence Interpreted as a Gateway from E-Cigarette Use to Smoking*

As early as 2011, it was speculated that e-cigarette use may be a 'gateway' to smoking combustible cigarettes,[1] mirroring prior concerns about smokeless tobacco and nicotine replacement therapies (NRTs).[2] Experimental study designs to establish causality are obviously unethical among such populations as non-smoking adolescents, so instead, longitudinal cohort (LC) data have been analyzed with LR-type models (including vanilla LR, repeated-measures generalized linear mixed models, etcetera) to investigate gateway effects among young people. More than a dozen such studies have been published, consistently reporting positive, statistically significant associations between prior e-cigarette use and subsequent smoking among adolescents (see e.g. [3]). An umbrella review concluded: "These repeated strong associations in prospective cohort studies are consistent with a causal relationship between vaping and subsequent smoking."[4]

*Evidence Against a Gateway from E-Cigarette Use to Smoking*

Population-level trends in prevalence of smoking and e-cigarette use among adolescents appear incompatible with a gateway effect. With a gateway, as e-cigarette use prevalence increases, the rate of decline of smoking prevalence would slow down unexpectedly (i.e., to a greater extent than otherwise expected if e-cigarettes did not exist) if the effect was weak, or even *reverse* (such that smoking prevalence actually increased) if the effect was strong. Numerous analyses of prevalence data from the US, UK, Germany, Aotearoa (New Zealand), and Taiwan using a range of nonlinear modelling techniques show that instead, there has been no unexpected slowing or reversal of the decline in smoking prevalence among adolescents as e-cigarette use prevalence increased, with some reporting a reverse-gateway or 'diversionary' effect [5-18] (see Appendix 1 for discussion of recent studies on Australia and Aotearoa [New Zealand]).

In adults, there is now high-certainty evidence from randomized trials that stop-smoking rates are higher for nicotine e-cigarettes than NRT.[19] Adults who switched from cigarettes to e-cigarettes are also less likely to relapse back to smoking than adults who used NRT,[20] and adults who 'dual use' cigarettes and e-cigarettes tend to reduce their cigarette consumption.[21-23] Additionally, econometric research has shown that cigarettes and e-cigarettes are economic substitutes, not complements.[24] These findings are not consistent with e-cigarettes leading to smoking; such an effect would seemingly imply lower stop-smoking rates, more relapse, and more cigarette consumption. It is inexplicable how the association between e-cigarette use and smoking would be opposite for adults and adolescents; we are not aware of any explanatory switch that flips on one's eighteenth (or twenty-first) birthday.

*The Apparent 'Paradox'*

The effects reported in LR-type studies of LC data (strong gateway effects with odds ratios up to 12)[3] and population-level trends (null or diversionary effects) are not compatible. This contradiction is illustrated in Figure 1 with a rudimentary simulation model (not intended for quantitative predictions; see Appendix 2 for details). What this simple model illustrates is that a weak gateway effect *slows* the rate of smoking decline, and a strong gateway effect *reverses* smoking decline. Importantly, neither of these scenarios are consistent with the actual, survey-measured smoking prevalence – hence the 'paradox', which requires investigation. Indeed, even *no* gateway effect (just the baseline historical decline) *over*-estimates actual smoking prevalence, suggesting that a diversionary effect is needed to explain population-level trends.

**Figure 1**. Trends in current (past 30-day) cigarette smoking and e-cigarette use among US middle and high school students from the National Youth Tobacco Survey (NYTS), compared to a rudimentary simulation model whereby smoking declines at the background rate circa 2009 but a proportion k of the increase in e-cigarette use is added to the smoking prevalence, counteracting the background decline (k=0.25, k=0.09, and k=0.00 for the 'Strong', 'Weak', and 'No' gateway effects, respectively). See Appendix 2 for details. Intended for illustration only; not quantitative forecasting.

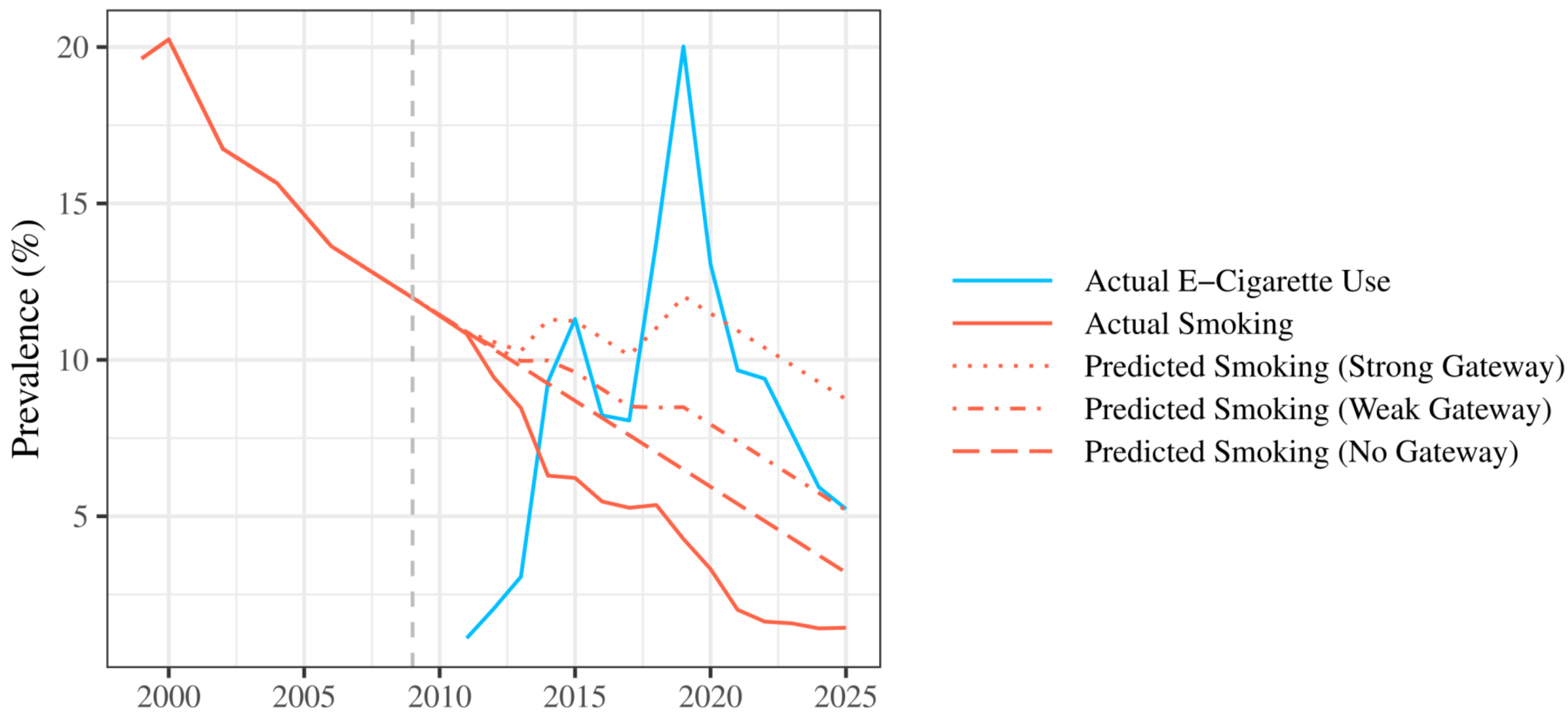

Since population-level trends are descriptive and represent reality, and since inferences from LR-type models should match what is observed descriptively at the population level, it is the LR-type models that are in question. Numerous criticisms of LR-type gateway studies have already been raised, including confounding and self-selection, low event counts, sub-cohort designs, and bi-directionality of associations (see Appendix 3 for details). The present study focuses instead on theoretical limitations of LR-type gateway models, how these preclude causal inference, and techniques available to overcome these limitations.

*Binary Response/Outcome*

Existing LR-type gateway studies model the response/outcome as a binary (two-level categorical) variable, e.g. current (past 30-day) smoking at follow-up (yes/no). But LC surveys used in such studies collect granular, numerical data on smoking behaviour, e.g. the number of days in the past 30 days on which a respondent smoked. Such data describe *quantitatively* the extent to which an adolescent is smoking; not at all (0 days), daily (30 days), and everything in-between. The binarization of numerical smoking data into two categories is problematic two-fold. The first is conceptual; collapsing numerical data down to a binary smoking indicator loses critical information. The difference between smoking on one vs 30 days at follow-up is highly clinically significant (experimental use vs likely nicotine dependence). Such differences are needlessly obscured by lossy compression of numerical data into two categories. The second is technical; binarization lowers statistical power and reliability, increases bias and inefficiency, introduces error, can introduce spurious effects (e.g. by increasing the risk of a positive result being a false positive), and is advised against by a substantial body of methodological literature (see [25] and citations therein). There is no theoretical justification for the ubiquity of the binarized smoking outcome; it is merely the road most travelled (despite severe limitations) and has become ingrained in the gateway literature.

*Other Limitations of Logistic Regression-Type Models*

The binarization issue described above could be addressed by fitting a Gaussian generalized linear model (GLM) with the (un-collapsed) numerical outcome rather than a binomial GLM (LR) with the binarized outcome. However, other limitations of linear models pertinent to causal inference remain, as described extensively in Appendix 4. The 'gist' is that logistic regression and related linear models are used by analysts as inference engines in the belief that the model will retrieve the independent influence of the predictors (e.g. baseline e-cigarette use) on the outcome (e.g. follow-up smoking), so that analysts can make inferences from the sample to the population (or in the case of prediction, the future population). But there is no guarantee that these models will do what most analysts believe they will do; what tends to happen instead is that these models *force* linear relationships onto the data with strong *a priori* assumptions about the existence and shape of said associations. Contrary to popular conception, these models do not in general provide comparative, independent effect estimates. As put bluntly by Carlin and Moreno-Bentacur,[26] "there are no reasonable assumptions under which the coefficients of a multivariable regression model simultaneously provide estimates of the causal effects of every variable in the model." This stems from the fact that LR-type models view the 'world' as a set of independent variables (selected by the analyst) all pointing independently (and linearly) toward (some function of) the outcome – but there is no guarantee that the world works this way, and these assumptions aren't tested by the models (or analysts). One of the consequences is that LR-type models struggle to recover nonlinear effects.

An important concept in causal inference is the ‘counterfactual’ scenario, namely what the outcome *would* have been for individuals if they had received a different treatment, which is what a causal effect estimate would describe (e.g., what follow-up smoking behaviour someone who used e-cigarettes at baseline *would* have had, if they had not used e-cigarettes). Odds/risk ratios from LR-type models do *not* describe such effects, cannot in general be considered in isolation from the model’s other variables, and cannot be readily interpreted as representing anything causal. Variance estimates from LR-type models may also be unreliable when there is insufficient common support in study variables across observations (which is common).

These severe limitations mean that LR-type models are basically never appropriate for causal inference, and are prone to ‘finding’ spurious associations. Regrettably, these limitations are overlooked in many basic statistics courses, with the result that LR-type models are thrown at problems for which they are generally inappropriate.

The extant literature on gateway effects in LCs relies on binarized outcomes and LR-type models. Consequently, tobacco control lacks studies using proper causal inference methods, having investigated gateway effects in a way that is severely limited at best. Attempts have been made to address these limitations within the LR-type model framework, e.g. by using propensity score matched (PSM) datasets to investigate gateway effects,[27] but unbeknownst to many analysts, PSM can actually *increase* imbalance, inefficiency, model dependence, and bias,[28] *worsening* the issues. Even if superior matching methods were used, LR-type models applied to matched data would still suffer the limitations described above.

Alternative models that address these limitations exist, but are at present extremely under-utilized. We introduce the unfamiliar reader to one such technique.

*BART: A Proper Causal Inference Technique*

A regression tree is a nonlinear and nonparametric approach to regression modelling by fitting decision trees to data. Vanilla regression trees can easily overfit, overemphasize multi-way interactions, do not provide direct uncertainty estimates, and depart from the true response surface when observations are scarce.[29] Bayesian Additive Regression Trees (BART) is a state-of-the-art, sum-of-trees model that addresses these limitations and those of LR-type models.[29, 30]

By building trees, BART does not force associations, but selects variables for inclusion in a principled, quantitative way, and handles interaction effects automatically (ignored in LR-type gateway studies). By lifting the assumption of linearity, BART is extremely flexible and can correctly model nonlinear relationships. By taking an ensemble approach (using multiple trees) and embedding this in a likelihood framework with prior specifications informed by the data themselves, BART controls model complexity and avoids overfitting. By working with counterfactual versions of the original dataset, BART provides proper causal effect estimates that go beyond simple odds/risk ratios. By taking a Bayesian approach, BART provides principled, reliable uncertainty estimates, namely Bayesian credible intervals, which estimate greater uncertainty where observations are scarcer (i.e., once the range of common support is left). The shift from the frequentist paradigm in previous LR-type gateway studies to the Bayesian paradigm in BART is nontrivial; whereas frequentist hypothesis testing asks how unusual are the data if the null hypothesis is true, Bayesian hypothesis testing asks how probable is the null hypothesis given the data. The two questions are in some sense opposites. Bayesian credible intervals are based on the observed data's likelihood, directly quantifying uncertainty about the causal effect estimate given the observations.

*Comparing LR-Type Models and BART*

Hoffman et al.[31] defines a step-by-step guide to causal study design using observational data. In brief, the steps relevant to the present discussion are: (1) Defining a research question (e.g., does e-cigarette use increase subsequent smoking?); (2) Defining an estimand or causal effect (e.g., the effect that e-cigarette use has on smoking at follow-up); and (3) Identifying and addressing potential sources of bias (e.g. confounding from shared smoking/vaping risk factors). Table 1 compares LR-type models to BART in terms of the limitations described above and these causal inference steps. In short, BART does what LR-type models don't. These improvements are not merely theoretical; BART has been shown empirically to out-perform LR-type models,[32] and BART won 1st place in the 2016 Atlantic Causal Inference Conference Black Box Competition;[33] in the same contest, the linear model placed just 13th. BART is not preferred arbitrarily, but because it is a better inference engine than LR-type models.

*Resolving the 'Paradox'*

The present study applies both LR and BART to the same longitudinal cohort data that have been analyzed in previous gateway studies, and shows that in contrast to LR-type studies and consistent with population-level trends, BART finds evidence *against* a causal gateway effect from e-cigarette use to smoking.

**Table 1:** Comparison of Logistic Regression-Type Models and Bayesian Additive Regression Trees

| | **Logistic Regression (LR)-Type Models** | **Bayesian Additive Regression Trees (BART)** |
|---|---|---|
| **Limitation** | | |
| Assumes that associations between the response and all explanatory variables exist *a priori*? | Yes; the linear predictor implicitly assumes *a priori* that associations exist between the response and all explanatory variables, represented by the coefficients/parameters $\boldsymbol{\beta}$. With this assumption, these models determine what the strength and direction (magnitude and sign of $\boldsymbol{\beta}$) for these associations *would* be *if* the model correctly represented reality, even if it does not. | No; BART's trees are built iteratively and select variables for inclusion from the explanatory variables provided based on quantitative selection criteria, which means *not* assuming *a priori* that associations exist between the response and all explanatory variables. |
| Assumes that associations between (some function of) the response and explanatory variables are linear, even though they may not be linear? | Yes; the linear combination of parameters and predictors in the linear predictor of the model equation means that the associations are assumed to be linear. The model does not test whether this assumption of linearity is correct. No automatic interaction handling is performed. | No; BART's trees are extremely flexible, allowing nonlinear relationships to be recovered. BART also handles interactions automatically. |
| Does not provide causal effect estimates, i.e. those based on counterfactual scenarios? | Yes; provides only odds/risk ratios that describe the odds/risk among one type of respondent compared to another type of respondent, which are prone to selection bias and do not represent causal effect estimates. | No; provides average treatment effects based on counterfactual versions of the dataset, which are proper causal effect estimates. |
| Estimates uncertainty unreliably when the covariate space is not well represented by all groups of a categorical variable? | Yes; performs no special adjustments to increase uncertainty when the covariate space is poorly represented across groups. | No; uncertainty estimates increase precipitously when the range of common support is left, providing reliable uncertainty estimation. The bartCause R package also natively provides common support rules to exclude observations lacking in common support. |

| | **Logistic Regression (LR)-Type Models** | **Bayesian Additive Regression Trees (BART)** |
|---|---|---|
| **Causal Framework Step** | | |
| Defining a research question | Does not intrinsically answer questions about e.g. what would have happened at follow-up to adolescents who did not use e-cigarettes at baseline had they done so (counterfactuals). | Intrinsically answers questions about counterfactuals by fitting the model to the original, unaltered LC dataset, predicting the smoking outcome at follow-up on counterfactual versions of the data (one in which all respondents are assigned to the treatment and one in which all respondents are assigned to the control) to generate posterior predictive distributions for either case, and finally taking the difference between these predictions for each respondent to generate a posterior distribution of the treatment effect.[29] |
| Defining an estimand or causal effect | Simply compares the odds/risk of smoking at follow-up between two groups who did and did not use e-cigarettes at baseline. These groups consist of different individuals, so the effect estimate is not isolated. | Generates a causal effect estimate by averaging the individual-level effects across the posterior distribution of the causal effect for the sample. The mean of the posterior distribution of the causal effect is the average treatment effect (ATE), a proper causal effect estimate.[29] Intuitively, this provides an estimate of what *would* have happened to someone had they or had they not used e-cigarettes at baseline, rather than crudely comparing two different groups as in LR-type models. |
| Identifying and addressing potential sources of bias | Adjustment for confounding by adding covariates to the model does not guarantee isolation of the effect of e-cigarette use because "there are no reasonable assumptions under which the coefficients of a multivariable regression model simultaneously provide estimates of the causal effects of every variable in the model."[26] | Can still suffer from confounding if insufficient confounders are provided, but isolates the effect of e-cigarette use better than LR-type models as it relaxes some of the latter's assumptions. The bartCause R package also natively provides methods such as propensity score as covariate, Targeted Minimum Loss based Estimation (TMLE) adjustment, and common support rules (which exclude observations lacking in common support) to reduce various sources of bias. Also, the Average Treatment Effect (ATE) and Average Treatment effect on the Treated (ATT) can be compared to identify selection bias. |

## METHODS

*Data*

Data were sourced from the Population Assessment of Tobacco and Health (PATH) Youth study, a nationally-representative (non-convenience-sample) LC study by the US Food and Drug Administration and National Institutes of Health, used in previous gateway studies with LR-type models (e.g. [34, 35]). PATH waves 3→4 were used since they cover the period 2015–2018 when e-cigarette use among US adolescents increased the most before peaking in 2019 and subsequently declining (see Figure 1 and [15]); if there is a gateway effect to be found in these data, it is most likely to be detected when e-cigarette use increased most, as that is when cigarette smoking is expected to increase most.

The analytic sample consisted of N=6,824 US adolescents aged 12–17 who had ever (n=466) or never (n=6,358) smoked combustible cigarettes at baseline (Wave 3), and who had non-missing data on the following baseline variables: number of days smoked cigarettes in past 30-days (numerical, 0–30 days), current (past 30-day) e-cigarette use (yes/no), ever use of other tobacco products (cigar, pipe, hookah, snus, smokeless, bidi, kretek, dissolvable; yes/no), susceptibility to smoke (yes/no; only defined for never-smoking adolescents; see [35]), sex (male/female), race (White only/Black only/other race), age (12–14/15–17), school performance (mostly B's and above/lower), family tobacco use (living with someone who uses tobacco; yes/no), peer tobacco use (has best friends who use tobacco; yes/no), secondhand smoke exposure (close contact with someone smoking in past seven days; yes/no), parental education (high school or GED or less/some college/college and higher), annual parental income (<\$50k/\$50k–\$99.999k/≥\$100k; unfortunately, continuous income data are not available in the public-use PATH datasets), current alcohol use (yes/no), and current cannabis use (yes/no); as well as follow-up (Wave 4) data on number of days smoked cigarettes in past 30-days (numerical, 0–30 days) and current (past 30-day) cigarette smoking (yes/no). The selection of and definitions for these variables were based on those by Sun et al.[35] and the recoded variables in the PATH public-use datasets, which are freely available from the Inter-university Consortium for Political and Social Research (https://doi.org/10.3886/ICPSR36498.v23). These analyses include 13 covariates covering socio-demographic, intra-individual, behavioural, peer influence, and family background factors in PATH associated with both smoking and e-cigarette use,[36] which is more covariates than almost all prior gateway studies have considered.[3, 36]

To address the binarization problem and leverage BART's support for numerical outcomes, a new variable, Δ days currently smoking, was defined as the number of days smoked in past 30-days at follow-up minus the number of days smoked in past 30-days at baseline (ranging from −30 [smoked on 30/30 days at baseline and 0/30 days at follow-up] to +30 [smoked on 0/30 days at baseline and 30/30 days at follow-up]).

*BART Model*

BART was implemented with the bartCause R package[29, 32] as a Gibbs sampler (a Markov Chain Monte Carlo [MCMC] method) with four chains, 1,000 posterior samples per chain (4,000 samples total), and a burn-in period of 1,500 samples to ensure sufficient mixing and convergence, assessed with trace plots. Δ days currently smoking was specified as the response, baseline e-cigarette use the treatment, and all other covariates described above were specified as the confounders. For confounders with >2 categories (e.g. parental income), one-hot encoding was used to produce dummy variables which retain all the same information. PATH survey weights were included in the model to account for the complex survey design, rescaled to sum to the unweighted sample size (N=6,824) while preserving relative magnitudes. Both treatment assignment mechanism and response surface were fitted with BART, and treatment effect estimates were computed with the doubly-robust 'tmle' method, which computes treatment effects using a propensity score-weighted sum of individual effects (with propensity scores calculated by bartCause using the confounders) and further adjusts individual effect estimates with the Targeted Minimum Loss based Estimation (TMLE) adjustment.[37] Propensity scores were also included as a covariate to help balance the groups and guard against regularization-induced confounding.[29, 38] Baseline smoking status (ever-smoking or never-smoking) was included in bartCause as a group factor to calculate the treatment effect estimate within each group from the same model.

Common support rules were used to improve common support and further reduce bias. As a sensitivity test, two common support rules were applied separately; the less stringent SD rule (cut-off=1 SD) and the more stringent chi-squared rule (cut-off *p*-value=0.05). See [39] for details.

*Model Fit*

Goodness-of-fit was assessed with Root Mean Square Error (RMSE; lower is better) for its interpretability, since it has the same units as the outcome (days). While this is technically a measure of predictive performance rather than causal inference, it is informative of model fit since BART performs prediction on the counterfactual datasets as part of its causal inference methodology (see Table 1).

*Significance of Results*

Statistical significance was assessed with Bayesian 95% highest density credible intervals. To improve clinical interpretation of findings, treatment effects were compared to a Minimal Clinically Important Difference (MCID). Clinical significance is context-specific and subjective, with no consensus on best method(s) to estimate/calculate MCID. Causal effect estimates for Δ days currently smoking were considered clinically significant if the change in the number of days smoking was ≥1 day in either direction (increase or decrease), since it is not clinically meaningful for an individual to smoke on e.g. 0.3 days (a day is smoked on or not).

*Linear Models*

As a replication of previous LR-type gateway studies, an LR model was fitted to the same PATH data with the same covariates using the survey R package[40] following PATH User

Guide guidance to account for the complex survey design. To accurately replicate previous studies, the outcome for the LR model was follow-up current cigarette smoking status (binary). We emphasize again that coding smoking behaviour as a binary variable as in previous LR-type studies needlessly discards *critical* information about smoking behaviour that is retained by the $\Delta$ days currently smoking variable in the BART model. For this reason, the binary outcome was not used in the BART model, and was used in the LR model strictly to demonstrate empirically how previous LR-type studies have produced spurious associations at odds with BART and population-level trends. Additionally, a Gaussian GLM was fitted to the $\Delta$ days currently smoking outcome, the purpose of which is to be verified by BART (but which *alone* is not a reliable inference engine).

## RESULTS

### *Sample*

Figure 2 shows the distribution of the sample (never- and ever-smoking respondents combined) across the study variables. Follow-up smoking and baseline use of e-cigarettes, other tobacco, alcohol, and cannabis were relatively rare among these adolescents, with survey-weighted prevalence estimates of 3.6%, 2.8%, 7.6%, 6.9%, and 3.0% respectively. <⅓rd of never-smoking respondents were susceptible to smoking. The sample was majority White only (69.9%) and 12–14 years old (60.6%), and roughly even between males and females.

**Figure 2**. Sample distribution for study variables. The outcome/response is shown in black, and the focal variable/treatment is in white. Following Lee at al.'s groupings,[36] behavioural factors are in red, intra-individual factors are in orange, peer influence factors are in yellow, socio-demographic factors are in green, and family background factors are in blue. *Susceptibility to smoke is only defined for never-smoking respondents.

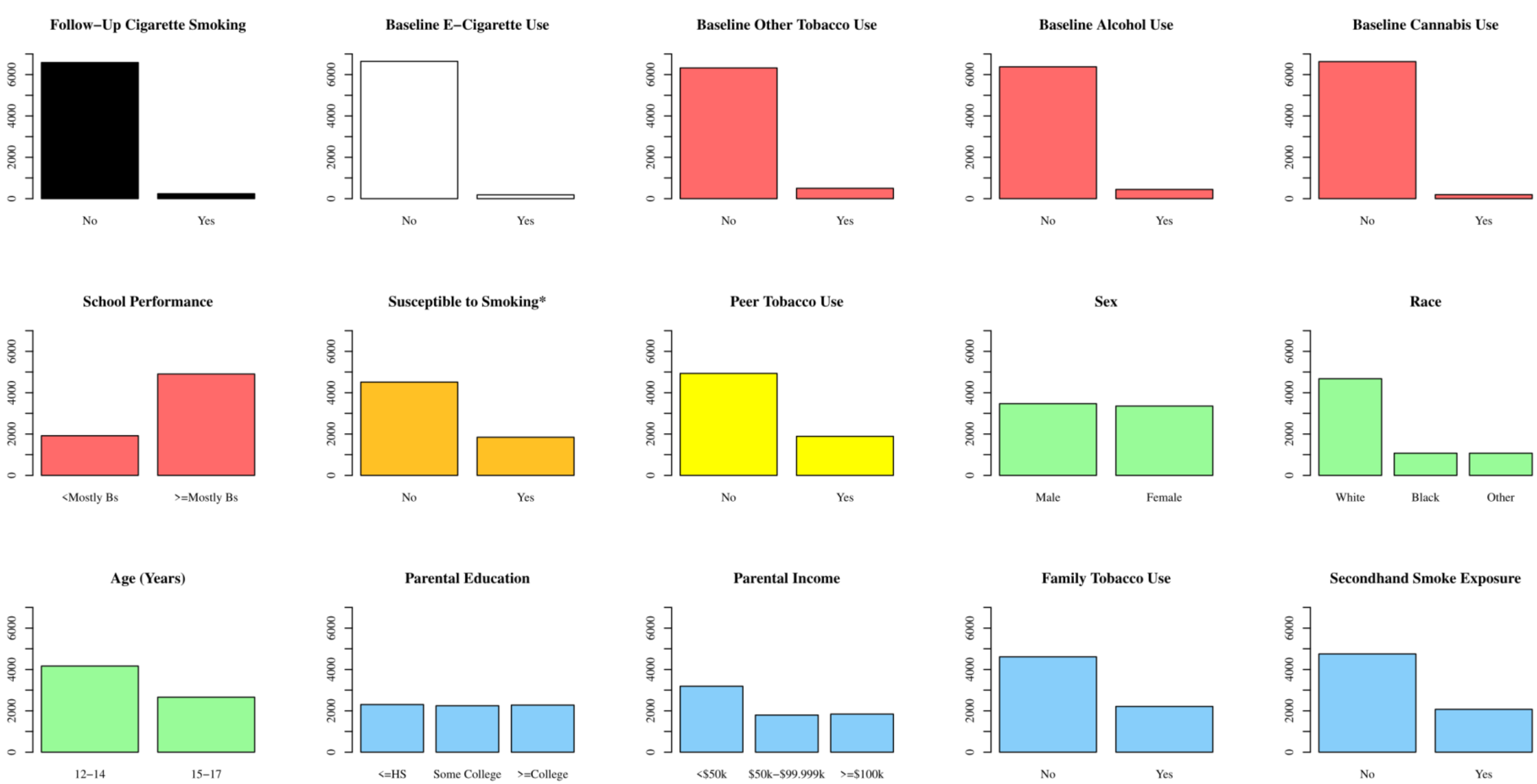

*BART Model Diagnostics*

Appendix 5 shows the model trace plots, exhibiting mixing and convergence, i.e., fluctuation around relatively stable means with rapid movement across parameter space, no long/flat patterns or long-term drifting, and over-lapping chains. In addition, the model RMSE was 1.6 days (for both common support rules), which is low relative to the range of the numerical outcome (60 days).

*BART Model Results*

BART model results are shown in Table 2. Across common support rules, the average treatment effect (ATE) of baseline e-cigarette use on $\Delta$ days currently smoking among never-smoking adolescents was 0.2–0.3 days, meaning that e-cigarette use was associated with an 'increase' in smoking days by ~¼th of a single day. This effect was not clinically significant (< the MCID of 1 day) and barely reached statistical significance. Oppositely, the ATE of baseline e-cigarette use on $\Delta$ days currently smoking among ever-smoking adolescents was −1.7 – −1.8 days, meaning that e-cigarette use was associated with a *decrease* in smoking days by approximately 2 days. This effect was both clinically significant (> the MCID of 1 day) and statistically significant. ATEs and Average Treatment effects on the Treated (ATTs; the average across only those respondents who used e-cigarettes at baseline) were not dissimilar, which may suggest low selection bias (in a randomized trial, ATE=ATT) and sufficient common support (though it does not rule out confounding). Appendix 6 shows waterfall plots for the individual treatment effects; diversionary effects (treatment effects<0 days) were both more common and generally stronger (larger in magnitude) than gateway effects (treatment effects>0 days).

*Linear Model Results*

By way of comparison, the LR model produce an 'adjusted' odds ratio (aOR) for baseline e-cigarette use of 1.9 (95% frequentist confidence interval: 0.7–5.1; Table 2) among the n=6,358 never-smoking respondents and 1.6 (1.0–2.6) among the n=466 ever-smoking respondents. Statistical significance aside, the interpretation of such effects in previous LR-type gateway studies is that e-cigarette use roughly doubles the odds of smoking (assuming that these ORs are correct causal estimates despite limitations described above). This gives the mistaken impression of a large gateway effect that is contradicted by both population-level trends (see Figure 1 and [5-18]) and the BART results.

Results from the Gaussian GLM were similar to BART in this particular case (Table 2), though confidence intervals for the linear model were wider. We emphasize strongly that the important result here is not that the Gaussian GLM captured effects resembling the BART effects, but that BART was necessary to verify the effects from the Gaussian GLM; the GLM's findings alone could not be considered causal. Again, the point is not that BART wasn't necessary, it's that it *was*. Indeed, synthetic empirical examples have shown that when linear model- and BART-recovered effects do differ, the BART effects are more accurate.[32] That the diversionary (ever-smoking group) and null (never-smoking group) effects here are model-independent (i.e., consistent across BART and the Gaussian GLM) for the numerical outcome is supportive of the validity of these findings.

**Table 2:** Model results for the effect of baseline e-cigarette use on follow-up smoking outcomes.

| | Baseline Never-Smoking Respondents (n=6,358) | Baseline Ever-Smoking Respondents (n=466) |
|---|---|---|
| BART Model (Outcome: Numerical*) | | |
| SD common support rule: | | |
| n suppressed** | 0 treatment, 3 control | 2 treatment, 9 control |
| ATE (95% Credible Interval) | 0.3 (0.1–0.5) | −1.7 (−2.4 – −1.1) |
| ATT (95% Credible Interval) | 0.4 (0.2—0.6) | −3.3 (−3.9 – −2.8) |
| Chi-squared common support rule: | | |
| n suppressed** | 0 treatment, 5,932 control | 0 treatment, 17 control |
| ATE (95% Credible Interval) | 0.2 (0.0–0.4) | −1.8 (−2.4 – −1.2) |
| ATT (95% Credible Interval) | 0.3 (0.0–0.5) | −3.3 (−3.9 – −2.8) |
| Gaussian GLM (Outcome: Numerical*) | | |
| Coefficient (95% Confidence Interval) | 0.3 (−0.3 – 1.0) | −2.1 (−3.9 – −0.3) |
| LR Model (Outcome: Binary***) | | |
| aOR (95% Confidence Interval) | 1.9 (0.7–5.1) | 1.6 (1.0–2.6) |

* Δ days currently smoking, defined as the number of days smoked in the past 30-days at follow-up minus the number of days smoked in the past 30-days at baseline.
** Number of observations suppressed by the common support rule to ensure common support and reduce bias.
*** Current (past 30-day) smoking status at follow-up (yes/no).

## DISCUSSION

This study explained the limitations of logistic regression-type models that preclude causal inference, and introduced the unfamiliar reader to BART; a model that addresses those limitations. The purported 'gateway' effect from e-cigarette use to smoking among adolescents, an apparent effect reported from LR-type models, was taken as an example and the effect was shown to disappear when BART is applied; no effect was found among never-smoking adolescents (consistent with 'common liability'; see Appendix 3), and a diversionary effect was found among ever-smoking adolescents, consistent with both prevalence trends (showing no unexpected increase or slowing of the decrease in smoking among adolescents[5-18]) and *adult* data from randomized trials and econometrics research.[19-24] These findings may resolve the apparent 'paradox' between LR-type gateway studies and population-level trends, as well as the apparent contradiction between studies in adolescents and adults. To the best of our knowledge, this is only the second time BART models have been applied in the context of e-cigarette use; recently, Xu et al.[41] applied BART to the PATH Adult study and found that e-cigarette use among adults who smoke was associated with a subsequent decline in smoking, similar to the present study findings.

That the results of the present study from an LC are consistent with studies of population level trends is supportive of their validity. Indeed, Pesko et al.[42] concluded that "[s]urvey data show youth cigarette use declining steadily despite vaping increasing. When past-30-day youth e-cigarette use rates were as high as 32.9% in 2019, youth smoking rates should have been rising if the [Surgeon General]'s statement that 'e-cigarette use is associated with the use of other tobacco products' represents a causal relationship. Instead, by 2021 the youth cigarette use rate fell to a record low 1.9%". It is perhaps worth noting that e-cigarette use is not the only hypothesised 'gateway' in substance use research. For years, cannabis was hypothesised to act as a 'gateway' to 'hard' drugs including heroin and cocaine. In their 2018 review, the National Institute of Justice concluded that "[t]he existing statistical research and analysis show mixed results and do not clearly demonstrate scientific support for cannabis use leading to harder illicit drug use. As a result, [the Federal Research Division] has determined that no causal link between cannabis use and the use of other illicit drugs can be claimed at this time."[43] That report is worthwhile reading for tobacco control researchers, because its criticisms are broadly applicable to the e-cigarette 'gateway' literature.

Comparisons can be made between the findings of the present study and previous studies. Whereas BART identified a clinically- and statistically-significant diversionary effect among ever-smoking adolescents and no effect among never-smoking adolescents, in a simulation model of the impacts of e-cigarette use on public health, Soneji et al.[44] assumed a very large gateway effect of +13.4 pp on the probability of smoking. In another modelling study, Harrell et al.[45] estimated that e-cigarette use increased the number of adolescents who smoke by 1.66 million from 2014–2019. Those effects are at odds with proper causal inference findings and rigorous analyses of population-level trends; such major discrepancies cast serious doubt on the validity and accuracy of the Soneji et al. and Harrell et al. studies (see also [46, 47]).

While some previous studies purporting to demonstrate 'gateway' effects refer simply to "associations" rather than making direct causal claims, it is more common than not for causality to be implied in the Abstracts and Discussion sections of these papers, for example by stating that the findings 'have implications for the regulation of e-cigarettes', which would only follow if e-cigarette use actually increased smoking. Examples of causal claims include an umbrella review stating that LR-type evidence is "consistent with a causal relationship between vaping and subsequent smoking";[4] a meta-analysis of LR-type studies claiming that "[e]ver e-cigarette use among never smokers at baseline quadruples the odds of being a smoker at follow-up";[48] and the Australian Government Department of Health claiming that "e-cigarettes can lead to an increased uptake of smoking among young people".[49] The present study suggests that the above claims are misleading at best.

*Limitations*

While this work addresses many of the limitations of LR-type studies, it is still subject to some of the other limitations discussed in Appendix 3, including confounding and low event counts. Indeed, like LR models, BART assumes that all confounders are measured.[29] In practice, it is impossible to account for all confounding in any study design that is not a completely randomized experiment because not all confounders are measured or even known. However, the number of confounders included in the present study exceeds almost all published gateway studies, and the similarity of the ATE and ATT in the present study may suggest limited selection bias. Low event counts are still problematic, and result from e-cigarette use and smoking being rare among these adolescents; we emphasize that like *all* LC gateway studies, event counts in this work were limited.

*Conclusion*

We have shown empirically that when longitudinal cohort data pertaining to gateway effects from prior e-cigarette use to subsequent smoking are analyzed using proper causal inference techniques that do not suffer from many of the limitations of models in the existing literature, the effect disappears. More than a dozen previous studies used severely limited models and came to the opposite conclusion, providing an illusion of causality that is lifted when the data are modeled appropriately with nonparametric

Bayesian causal inference techniques. Proper causal inference techniques appear to reconcile the apparent contradiction between population-level trends and problematic analyses of longitudinal cohorts. Public health in general and tobacco control in particular are not in need of more logistic regression-type models, but research with more robust and rigorous computational techniques that are less likely to produce spurious associations, i.e., a paradigm shift from logistic regression (and related linear models) to Bayesian nonparametric modelling for causal inference.

**Data Availability** The Population Assessment of Tobacco and Health (PATH) public-use datasets supporting the conclusions of this article are freely available from the Inter-university Consortium for Political and Social Research, https://doi.org/10.3886/ICPSR36498.v23

**Funding** None.

**Declaration of Interests** This work received no funding and builds upon previous findings presented at the 2025 Annual Society for Research on Nicotine and Tobacco - Europe (SRNT-E) Conference in Cluj Napoca, Romania, September 10–12, 2025.[50] Author FF is a graduate student in Artificial Intelligence and Machine Learning at The University of Texas at Austin, which had no involvement in this work, and separately provides consulting services through Pinney Associates on tobacco harm reduction to Juul Labs Inc., which also had no involvement in this work. Pinney Associates has previously consulted to other commercial nicotine product manufacturers in the pharmaceutical and tobacco industries on US regulatory pathways for non-combusted and non-tobacco nicotine products, which also had no involvement in this work. Neither FF nor Pinney Associates consult on combustible tobacco products. Author RN is a Professor of Social and Behavioral Sciences at New York University and has received grant and contractual funding from the National Institutes of Health and the Food and Drug Administration; served as a paid consultant to the Government of Canada via a contract with Industrial Economics Inc; received an honorarium for a virtual meeting from Pfizer Inc.; received other NIDA grants paid to his employers; received salary from the Steven Schroeder Institute for Tobacco Research and Policy Studies at The Legacy Foundation, now Truth Initiative, New York University School of Global Public Health; and communicated with Juul Labs personnel, for which there was no compensation, and received hospitality in the form of meals at some meetings; none of which supported the work reported here. The Progressive Policy Institute sponsored a trip (travel and lodging) for RN to present a paper at a symposium: can e-cigarettes help tobacco cigarette smokers quit? A review of the evidence, Tobacco Harm Reduction—an Update, 54th annual meeting of the Japanese Society of Neuropsychopharmacology, jointly held with the 34th annual meeting of the Japanese Society of Clinical Neuropsychopharmacology and the 35th World Congress Collegium International Neuro-Psychopharmacologicum, Tokyo International Forum, Tokyo, Japan, May 24, 2024; no honorarium, consulting fee or other payment was provided.

**Author Contributions** Floe Foxon (Conceptualization, Data curation, Formal analysis, Investigation, Methodology, Software, Visualization, Writing – original draft), Raymond Niaura (Conceptualization, Investigation, Methodology, Writing – review & editing)

## Appendix 1: Discussion of other population-level trends literature

A pair of articles by Egger et al.[1, 2] are among the few population-level trend studies purporting to show evidence for a gateway effect from e-cigarette use to smoking among adolescents. Those analyses considered smoking prevalence data from the Australian Secondary Students' Alcohol and Drug Survey (ASSAD) from 1999 to 2022/3 and Aotearoa (New Zealand) Action on Smoking and Health (ASH) Year 10 Snapshot Survey from 1999 to 2023.

Following previous work,[3] the Egger et al. analyses fitted regression models to smoking prevalence through to some cut-off year. This 'background trend' can then be projected forward in time, acting as a counterfactual scenario (what would have happened to smoking prevalence if trends from before e-cigarettes continued), and compared to actual survey-measured smoking prevalence in the post-cut-off period. However, unlike previous work,[3] Egger et al. used generalized linear models, i.e. those with linear predictors (a linear combination of parameters and explanatory variables), and did not provide uncertainty bands around the projected trends. The decision to not illustrate uncertainty bands in the Egger et al. figures is inexplicable, since the models were not merely illustrative but were intended for quantitative prediction of future scenarios (forecasting), which carries considerable uncertainty. Additionally, the assumption of a linear predictor may be unduly restrictive. Rather than using a generalized linear model with a linear predictor, non-linear regression could be used (as in [3]), e.g. an exponential decay curve of the form: $\text{smoking prevalence}_{\text{year}} = \alpha e^{-\beta \times \text{year}}$, which has no more parameters than the logistic regression model; which, like the logistic regression model, cannot produce negative prevalence predictions; and which fits the data equally well (as measured e.g. by Root Mean Square Error [RMSE]).

When the non-linear model is fitted to the ASSAD and ASH past-month smoking prevalence data (including 1996 ASSAD data and the latest 2024 and 2025 ASH data not used by Egger et al.) using the nls (Nonlinear Least Squares) function in R, actual/observed past-month smoking prevalence in the post-cut-off period is consistent with the 95% prediction interval of the counterfactual/ background trends, which were estimated with bootstrapping (N=10,000 iterations) using the nlsBoot and nlsBootPredict functions in the nlstools R package (see figure below). In other words, smoking prevalence was no greater than expected. This directly counters the Egger et al. claim that "the rapid rise of vaping may have slowed the rates of [smoking] decline"; instead, smoking declines in Aotearoa (New Zealand) and Australia appear to have continued as usual with no gateway effect.

Besides these logistic and nonlinear models, other valid and equally justifiable models with other conclusions could almost certainly be found. Gary King refers to this as "model dependence", whereby models of the same data with different (but valid and justifiable) functional forms result in different conclusions.[4] On studies with model dependence like those of Egger et al., King asks: "Is this a true test of an ex ante hypothesis or merely a demonstration that it is *possible* to find results consistent with your favorite hypothesis?" Given that no slowing of the rate of decline of smoking associated with e-cigarette use has been observed with recent data elsewhere, and given other evidence against the gateway effect cited in the Introduction section of the present study, Egger et al.'s conclusions appear to be the latter of King's cases. See [5] for further criticism of the Egger et al. analyses.

**Trends in past-month adolescent smoking prevalence in Aotearoa (New Zealand) and Australia.** Non-linear regression models fitted to smoking prevalence from 1996/9 to 2008 (the cut-off year used by Egger et al.[2]) and projected forward in time. Shaded areas represent 95% prediction intervals.

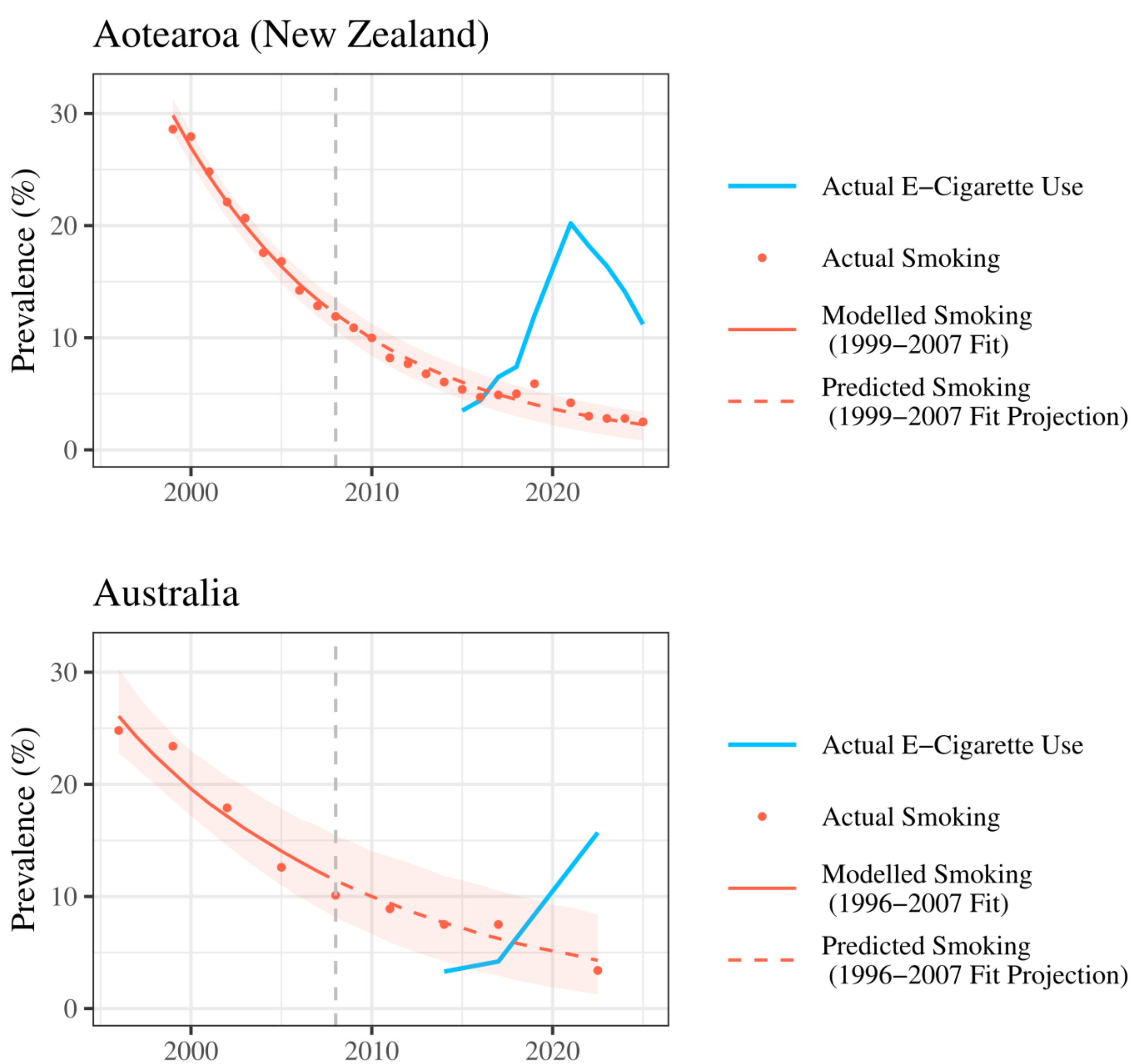

**Appendix 2: Simulation model details**

The rudimentary and illustrative simulation model compares actual, survey-measured smoking prevalence (current [past 30-day] use) among middle and high school students in the National Youth Tobacco Survey (NYTS) to three counterfactual scenarios:

1. 'No Gateway': Smoking continues to decline after 2009 (circa the initial uptake of e-cigarettes) at the same rate as it did in the few years up to 2009, with no added effect of e-cigarette use.
2. 'Weak Gateway': Same as (1) but with a weak gateway effect added on top of this background decline such that any increase in e-cigarette use between years increases the smoking prevalence by k=0.09 times the increase in e-cigarette use. Roughly speaking, 9% of the increase in e-cigarette use is added to the smoking prevalence. 9% was selected through trial and error as the maximum effect that would not actually increase year-over-year smoking prevalence (at most, smoking is flat year-over-year).
3. 'Strong Gateway': Same as (2) but with a stronger gateway effect with k=0.25. Roughly speaking, 25% of the increase in e-cigarette use is added to the smoking prevalence. 25% was selected arbitrarily as it corresponds to a quarter; k could be any value >0.09 under a strong gateway scenario.

Of course, a linear model projected into the future would eventually predict a *negative* smoking prevalence, which is not possible. This model is illustrative and is not used to make numerical predictions; as noted in Appendix 1, proper forecasting models require the quantification of uncertainty. Since the 2021 NYTS was administered roughly equally at-home (COVID) and in-school whereas all other years of NYTS were administered exclusively or almost-exclusively in-school, the 2021 point estimates in this work include in-school respondents only to improve comparability across years. NYTS data are publicly available from: https://www.cdc.gov/tobacco/about-data/surveys/historical-nyts-data-and-documentation.html and https://www.fda.gov/tobacco-products/youth-and-tobacco/national-youth-tobacco-survey-nyts. Since this model is not produced by standard R packages, R code is provided below:

```
## IMPORT PACAKGES ##
library(ggplot2) # import data plotting package

## DATA ##
# year
year = c(1999, 2000, 2002, 2004, 2006, 2009, 2011, 2012, 2013, 2014, 2015, 2016,
         2017, 2018, 2019, 2020, 2021, 2022, 2023, 2024, 2025)
# current (past 30-day) cigarette smoking prevalence (1999-2025)
cig = c(19.6324, 20.2450, 16.7436, 15.6463, 13.6336, 11.9860, 10.8277, 9.4409,
        8.4642, 6.2975, 6.2261, 5.4710, 5.2713, 5.3594, 4.2741, 3.3151, 2.0098,
        1.6320, 1.5762, 1.4155, 1.4356)
# current (past 30 day) e-cigarette use prevalence (2011-2025)
eCig = c(1.1042, 2.0472, 3.0660, 9.3020, 11.3053, 8.2246, 8.0604, 13.7767,
         20.0187, 13.0610, 9.6654, 9.3972, 7.6652, 5.9287, 5.2317)

## SIMULATION ##
# average annual decline in smoking prevalence 2006-2009
averageAnnualDecline = (13.6336-11.9860)/3

# LINEAR DECLINE + NO GATEWAY
# smoking prevalence 1999-2009
cigLinear = cig[1:6]
# for each year from 2010 to 2025...
for(x in seq(2010, 2025, 1)){
  # the next synthetic smoking prevalence is the previous year's prevalence
  # minus the average annual decline before 2010
  nextCig = tail(cigLinear, 1) - averageAnnualDecline
  # add to list of synthetic smoking prevalence
  cigLinear = append(cigLinear, nextCig)
}
```

```
# for plotting purposes, delete the 2010 estimate (for which there is no NYTS)
cigLinear = cigLinear[-7]

# LINEAR DECLINE + WEAK GATEWAY
# smoking prevalence 1999-2009
cigLinearGatewayWeak = cig[1:6]
# gateway constant (proportion of new e-cig-using respondents who go on to smoke)
k = 0.09
# year-over-year change in e-cigarette use
ecigDiffs = diff(eCig)
# no NYTS data for 2009 and 2010 e-cig use; assume negligible (<1% prevalence)
ecigDiffs = append(0, ecigDiffs)
ecigDiffs = append(0, ecigDiffs)
# for each year from 2010 to 2025...
for(i in seq(1, 16, 1)){
  nextCig = 0
  # if e-cigarette use increased year-over-year...
  if(ecigDiffs[i] > 0){
    # the next synthetic smoking prevalence is the previous year's prevalence
    # minus the average annual decline before 2010 plus gateway effect
    nextCig = tail(cigLinearGatewayWeak, 1) - averageAnnualDecline +
      k*ecigDiffs[i]
  }else{
    # the next synthetic smoking prevalence is the previous year's prevalence
    # minus the average annual decline before 2010 (no gateway effect)
    nextCig = tail(cigLinearGatewayWeak, 1) - averageAnnualDecline
  }
  # add to list of synthetic smoking prevalence
  cigLinearGatewayWeak = append(cigLinearGatewayWeak, nextCig)
}
# for plotting purposes, delete the 2010 estimate (for which there is no NYTS)
cigLinearGatewayWeak = cigLinearGatewayWeak[-7]

# LINEAR DECLINE + STRONG GATEWAY
# smoking prevalence 1999-2009
cigLinearGatewayStrong = cig[1:6]
# gateway constant (proportion of new e-cig-using respondents who go on to smoke)
k = 0.25
```

```
# year-over-year change in e-cigarette use
ecigDiffs = diff(eCig)
# no NYTS data for 2009 and 2010 e-cig use; assume negligible (<1% prevalence)
ecigDiffs = append(0, ecigDiffs)
ecigDiffs = append(0, ecigDiffs)
# for each year from 2010 to 2025...
for(i in seq(1, 16, 1)){
  nextCig = 0
  # if e-cigarette use increased year-over-year...
  if(ecigDiffs[i] > 0){
    # the next synthetic smoking prevalence is the previous year's prevalence
    # minus the average annual decline before 2010 plus gateway effect
    nextCig = tail(cigLinearGatewayStrong, 1) - averageAnnualDecline +
      k*ecigDiffs[i]
  }else{
    # the next synthetic smoking prevalence is the previous year's prevalence
    # minus the average annual decline before 2010 (no gateway effect)
    nextCig = tail(cigLinearGatewayStrong, 1) - averageAnnualDecline
  }
  # add to list of synthetic smoking prevalence
  cigLinearGatewayStrong = append(cigLinearGatewayStrong, nextCig)
}
# for plotting purposes, delete the 2010 estimate (for which there is no NYTS)
cigLinearGatewayStrong = cigLinearGatewayStrong[-7]

## PLOT
# plot results
ggplot() +
  geom_line(aes(x = tail(year, -6), y = eCig, color = 'Actual E-Cigarette Use',
               linetype = 'Actual E-Cigarette Use')) +
  geom_line(aes(x = year, y = cig, color = 'Actual Smoking',
               linetype = 'Actual Smoking')) +
  geom_line(aes(x = year, y = cigLinearGatewayWeak, color = 'Predicted Smoking
(Weak Gateway)',
               linetype = 'Predicted Smoking (Weak Gateway)')) +
  geom_line(aes(x = year, y = cigLinearGatewayStrong, color = 'Predicted Smoking
(Strong Gateway)',
               linetype = 'Predicted Smoking (Strong Gateway)')) +
```

```
geom_line(aes(x = year, y = cigLinear, color = 'Predicted Smoking (No Gateway)',
          linetype = 'Predicted Smoking (No Gateway)')) +
theme_bw(base_family = 'serif', base_size = 16) +
theme(legend.key.width = unit(3.5, "line")) +
labs(x = '', y = 'Prevalence (%)') +
scale_color_manual(name = '',
             breaks = c('Actual E-Cigarette Use',
                     'Actual Smoking',
                     'Predicted Smoking (Strong Gateway)',
                     'Predicted Smoking (Weak Gateway)',
                     'Predicted Smoking (No Gateway)'),
             values = c('Actual E-Cigarette Use' = 'deepskyblue',
                     'Actual Smoking' = 'tomato',
                     'Predicted Smoking (Strong Gateway)' = 'tomato',
                     'Predicted Smoking (Weak Gateway)' = 'tomato',
                     'Predicted Smoking (No Gateway)' = 'tomato')
             ) +
scale_linetype_manual(name = '',
              breaks = c('Actual E-Cigarette Use',
                      'Actual Smoking',
                      'Predicted Smoking (Strong Gateway)',
                      'Predicted Smoking (Weak Gateway)',
                      'Predicted Smoking (No Gateway)'),
              values = c('Actual E-Cigarette Use' = 'solid',
                      'Actual Smoking' = 'solid',
                      'Predicted Smoking (Strong Gateway)' = 'dotted',
                      'Predicted Smoking (Weak Gateway)' = 'dotdash',
                      'Predicted Smoking (No Gateway)' = 'longdash')
              ) +
geom_vline(xintercept = 2009, linetype = 'dashed', col = c('gray'))
```

## Appendix 3: Existing criticisms of gateway studies in longitudinal cohorts

| Criticism | Explanation |
|---|---|
| Confounding and self-selection | One of the most notable issues with prior LR-type studies is insufficient adjustment for confounding from common liabilities or shared risk factors associated with both vaping and smoking. One commentator has claimed that criticisms of LR-type studies do not specify which confounding variables are unadjusted for and how this inadequate adjustment for confounding could account for observed associations.[6] But in fact, Lee et al.[7] have identified numerous factors related to use of either product, of which individual LR-type studies typically include ≤10 (see Supplemental Materials of Lee et al.); and Chan et al.[8] have shown empirically that published estimates for gateway effects have low E-values compared to known risk factors, and are, therefore, not robust against unmeasured confounding. Lee et al.[9] have also shown empirically that the more confounders are included in LR-type models, the closer to the null value the effect estimate becomes. Thus, gateway confounders are real and materially affect results. As noted by Delnevo,[10] "youth tobacco-use behaviors are complex, and experimentation with multiple tobacco products is common [among those using nicotine]." Those adolescents who would experiment with one type of product are likely the same adolescents who would experiment with another type of related product. Thus, differences in odds of smoking initiation are likely due simply to differences in the groups of adolescents who experiment with substance use (self-selection), and not because one product causes use of the other. |
| Low event counts | The total analytic sample sizes in previously-published LR-type studies range from N=246 to N=10,384.[11] Because smoking is rare among adolescents (e.g., just 1.4% of US middle and high school students reported current smoking in 2025 [see Figure 1 of this study]), the number of respondents in LC datasets reporting e-cigarettes use but no smoking at baseline as well as smoking at follow-up (i.e., the number of 'events') ranges from just N=6 to N=124 total in previously-published LR-type studies.[11] Van der Ploeg et al.[12] have shown empirically that logistic regression, the type of linear model used almost exclusively in LC gateway studies, requires 20 to 50 events *per variable* to reach a stable area under the curve. |

| Criticism | Explanation |
| --- | --- |
| Sub-cohort design | By focusing on only those respondents who use e-cigarettes but do not smoke at baseline, it has been shown empirically that LR-type gateway studies exclude a large proportion of respondents who experiment with smoking at follow-up, namely those who report some experimentation with smoking at baseline (with or without e-cigarette use) and who are therefore relevant to investigating factors associated with smoking at follow-up.[13] In so doing, LR-type gateway studies preclude diversionary effects. |
| Bidirectionality | LR-type models not only produce significant, positive associations between prior e-cigarette use and subsequent smoking, but also significant, positive associations between prior smoking and subsequent e-cigarette use.[14] If these associations were causal, then this bi-directionality would imply an endless cycle of cause and effect that is both inexplicable and inconsistent with a net gateway. Instead, these bi-directional associations are more supportive of common liability. |

**Appendix 4: Detailed limitations of logistic regression-type models**

<u>*A Priori* Assumption of Associations</u>: LR-type models are generalized linear models. In simple terms, the form of the model equation in LR-type studies is:

$$\text{logit}(\text{Pr}(\text{Follow-up smoking})) = \beta_0 + \beta_1 \text{baseline vaping} + \beta_2 \text{confounder}_1 + \cdots + \beta_{m+1} \text{confounder}_m$$

where Pr(Follow-up smoking) is the probability of smoking at follow-up, $\beta_0$ is the intercept, and $\text{confounder}_i$ is the $i^{\text{th}}$ of m risk factors for smoking and vaping (e.g., age). Implicit in the model equation is the *a priori* assumption of possible associations between (a function of) the response and explanatory variables as 'natural laws' (which are forced), represented by the non-intercept coefficients/parameters $\beta$. With this assumption, LR-type models determine what the strength and direction (magnitude and sign of $\beta$) for these associations *would* be *if* the model correctly represented reality (see e.g. [15])…

But does the model correctly represent reality? There is no guarantee that it does. This is not addressed by *p*-values, which are often misinterpreted as the probability that an effect is 'real'.[16] Attempts to address this by excluding certain explanatory variables based on statistical significance should never be done, as it is without foundation and statistically incorrect[17] (despite also being ubiquitous). The issue is that LR-type models are parametric and do not select which variables are or are not included in the equation and may therefore be associated with the outcome on a well-justified quantitative basis; that decision is left to the analyst, and it is usually arbitrary. Logistic regression was not designed for variable selection.

<u>Assumption of Linear Associations</u>: The linear predictor (right-hand side of the equation above) is a *linear* combination of the parameters and explanatory variables, such that associations between (a function of) the response and the explanatory variables are assumed to be linear…

But are the associations linear? There is no good reason to assume that they are. LR-type models do not assume linearity because there is a good theoretical basis for doing so; it is merely a mathematical convenience[18] from a time when computation was limited, that has carried over into modern epidemiology because such models are relatively straightforward to teach students who become the researchers that use these models. LR-type models could only be causal models if the parametric model (i.e., the assumption of linear associations) is correct.[19] If the linear parametric model is not correct, e.g. because the 'true' associations are nonlinear, then the model *cannot* recover the causal effect of an explanatory variable on the outcome (e.g. of prior vaping on subsequent smoking), and any effect estimates from such models will be biased and unreliable.

Lack of Counterfactuals and Causal Estimates: After LR-type models are fitted to data, the parameters $\beta$ are exponentiated to produce odds ratios. These odds ratios are often interpreted as estimates of the causal effect of a given explanatory variable on the outcome with all other variables 'held' constant, e.g., the effect that baseline vaping has on follow-up smoking independent of the other explanatory variables or covariates…

But is this interpretation correct? The answer is no, and this is known as the 'Table 2 Fallacy'.[20] In the context of gateway studies, the odds ratio describes the ratio between the odds of follow-up smoking among the type of respondents who used e-cigarettes at baseline, and the odds of follow-up smoking among the type of respondents who did not used e-cigarettes at baseline, assuming that the association exists and is linear. The odds ratio is *not* a proper causal effect estimate. Even if the linearity assumption was correct, the odds ratio does *not* describe what the smoking outcome *would have been* for respondents who used e-cigarettes at baseline had they *not* done so, or what the smoking outcome *would have been* for respondents who did not use e-cigarettes at baseline had they done so, which is what proper causal effect estimates would describe.[21]

‘Counterfactual’ scenarios such as these are essential to investigating the causal effect of an explanatory variable on the outcome, and they are *not* estimated in LR-type models. Further, the odds ratio is not truly independent of other factors: as stated by Carlin and Moreno-Bentacur,[18] “there are no reasonable assumptions under which the coefficients of a multivariable regression model simultaneously provide estimates of the causal effects of every variable in the model.”

Unreliable Variance Estimation: After LR-type models are fitted to data and odds ratios are obtained, confidence intervals are used to determine the statistical significance of the effect…

Are these confidence intervals reliable? It depends. Even if the odds ratios from LR-type studies were true causal effect estimates, if the covariate space (i.e., the space represented by all possible values of all explanatory variables) is not well represented by both groups in the binary variable of interest/treatment (e.g. e-cigarette use), then variance will not be estimated reliably. E.g., if there is insufficient overlap or ‘common support’ between those who used e-cigarettes at baseline and those who did not in terms of age, sex/gender, race/ethnicity, and a whole host of other included confounders, then the estimate for the confidence interval around the e-cigarette use odds ratio will not be reliable. Note: this is a separate variance estimation issue to (multi)collinearity.

Besides causal inference, LR-type models may yet have good predictive performance even if the assumed linear associations are incorrect. Importantly however, in such cases the coefficients/parameters of the model $\beta$ are merely constants for use within a prediction algorithm and have no meaningful interpretation.[18] Worse still, published LR-type gateway models have been shown empirically to have low predictive validity,[11] so even this use case is limited in the context of the gateway effect.

**Appendix 5: Trace plots for BART models.** The blue chains correspond to treatment effects for the never-smoking sample. The red chains correspond to treatment effects for the ever-smoking sample. From top to bottom: Chi-squared common support rule, ATE and ATT; SD common support rule, ATE and ATT.

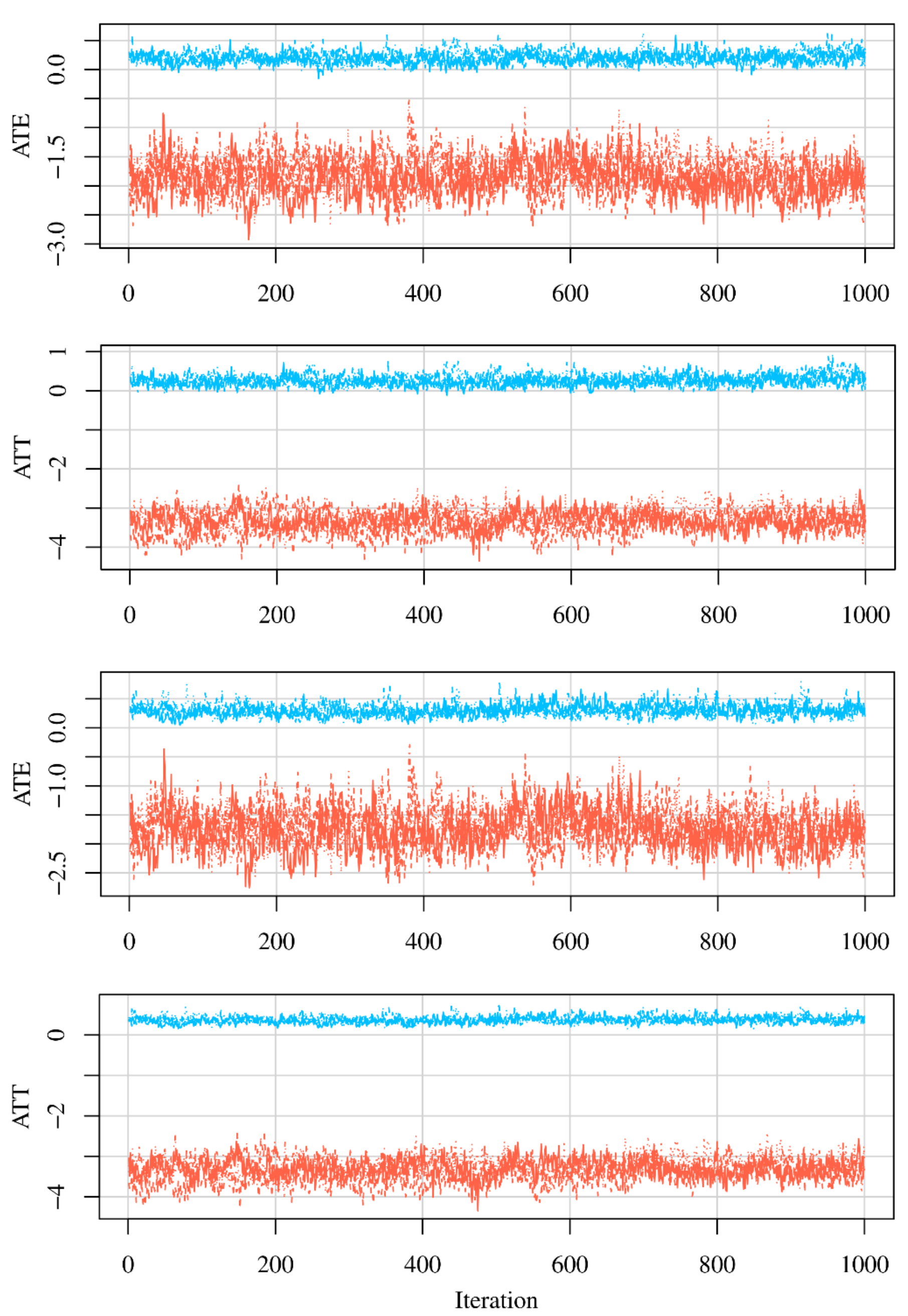

**Appendix 6: Waterfall plots of individual treatment effects in BART models.** Points and posterior intervals for individual effects are in green. The blue area represents the 95% credible interval of the ATE/T for the never-smoking sample. The red area represents the 95% credible interval of the ATE/T for the ever-smoking sample. From top to bottom: Chi-squared common support rule, ATE and ATT; SD common support rule, ATE and ATT.

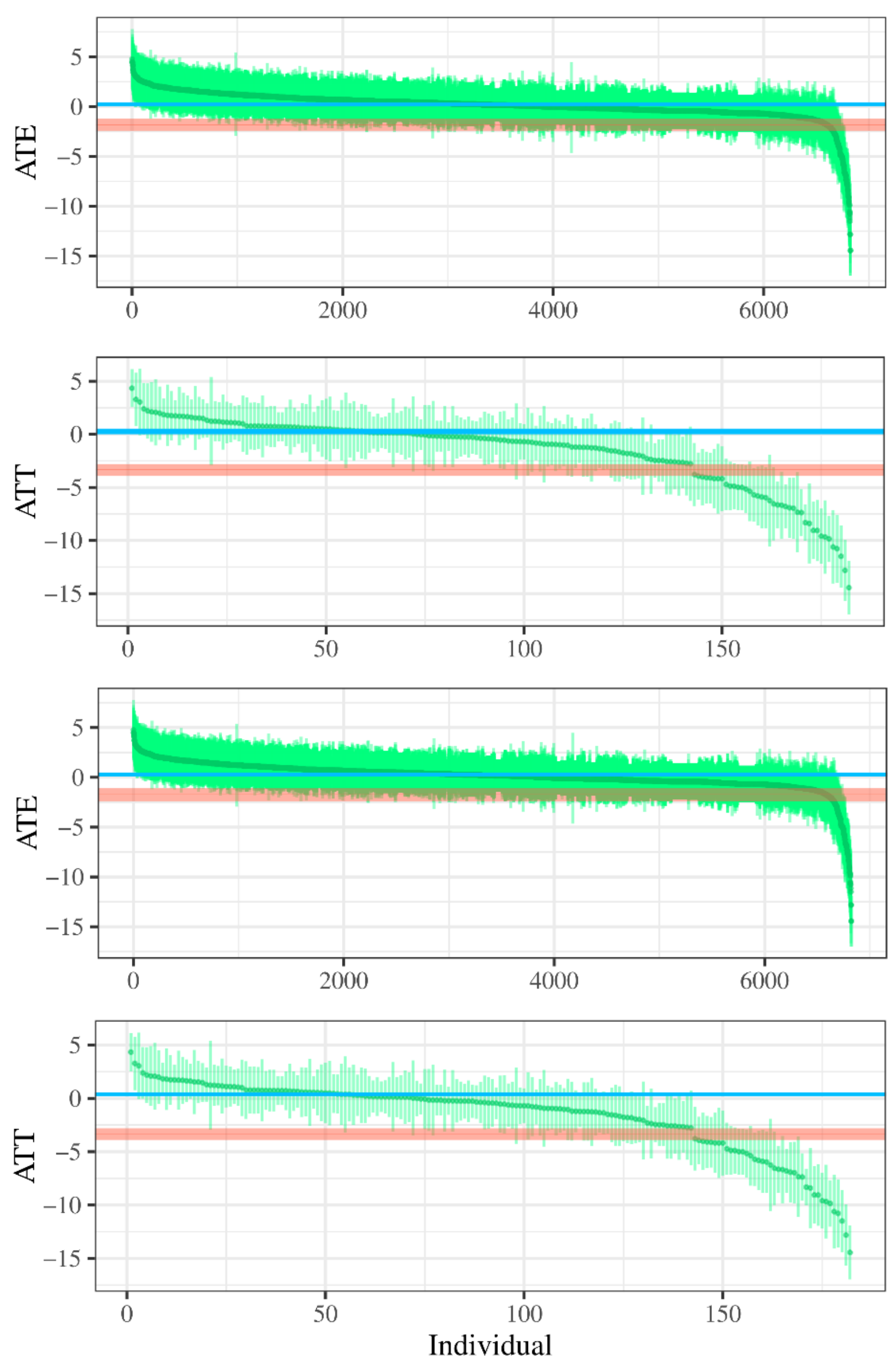

**Appendix References**